\documentclass[pop,fleqn]{w-art}
\usepackage{amsfonts}
\usepackage{amssymb}
\usepackage{times}
\usepackage{w-thm}
\usepackage{graphicx}
\usepackage{subfigure}

\newcommand{\googol}{\mathrm{Googol}}
\newcommand{\be}[1]{\begin{equation}\label{#1}}
\newcommand{\ee}{\end{equation}}
\newcommand{\bea}[1]{\begin{eqnarray}\label{#1}}
\newcommand{\eea}{\end{eqnarray}}

\newcommand{\bZ}{\mathbb{Z}}

\newcommand{\N}{\bf N}

\newcommand{\Anti}{{\bf Anti}}
\newcommand{\Sym}{{\bf Sym}}

\newcommand{\ov}{\overline}
\newcommand{\mtt}{\footnotesize\tt}
\newcommand{\ga}{\alpha}
\providecommand{\href}[2]{#2}
\begin{document}
\pagespan{1}{}
\bigskip

\begin{flushright}
\hfill{MPP-2005-164}\\
\hfill{LMU-ASC 80/05}\\
\hfill{hep-th/0512190}\\
~ 
\end{flushright}

\title[Standard Model statistics of a Type II orientifold]{Standard Model statistics of a Type II orientifold}

\author[F. Gmeiner]{Florian Gmeiner\inst{1,2,}%
  \footnote{\textsl{flo@mppmu.mpg.de}}}
\address[\inst{1}]{Max-Planck-Institut f\"ur Physik\\F\"ohringer Ring 6, 80805 M\"unchen, Germany}
\address[\inst{2}]{Arnold-Sommerfeld-Center for Theoretical Physics\\Department f\"ur Physik, Ludwig-Maximilians-Universit\"at  M\"unchen\\Theresienstra{\ss}e 37, 80333 M\"unchen, Germany}

\begin{abstract}
We analyze four-dimensional,
supersymmetric
intersecting D-brane models in a toroidal orientifold background
from a statistical perspective.
The
distribution and correlation of observables, like gauge groups and couplings,
are discussed.
We focus on models with a Standard Model-like gauge sector,
derive frequency distributions for their occurrence and analyze the
properties of the hidden sector.

\bigskip
\noindent\emph{Talk given at the RTN Workshop ''Constituents, Fundamental Forces and Symmetries of the Universe'', Corfu, Greece, 20-26 Sep 2005.}

\end{abstract}
\maketitle 

%\tableofcontents

\section{Introduction}\label{sec_intro}
The search for string vacua which could provide
the Standard Model gauge group
at low energy is an important aspect of
research in string theory in order to make contact
with phenomenology.
The models which have been considered most in the
literature are
compactifications of the heterotic string
and type II
orientifold constructions \cite{Blumenhagen:2005mu}.
The study of type II flux vacua \cite{Grana:2005jc}
has led to the conclusion
that there might exist a huge number ($\approx\googol^5$)
of possible stable vacua.
This prospect shows clearly that new methods which go beyond
explicit constructions of single models have to be
developed to deal with this huge number.

One idea that was put forward in \cite{Douglas:2003um} and
further developed by several authors, suggests to change the
analysis from a model-by-model construction to a statistical
approach. If one were able to detect some general behavior
in the statistics, independent of the specific model
chosen, this might exclude large regimes of the
landscape of vacua and also give us some hints where to
look for realistic models.
So far most of the statistical analysis has been done in the
closed string sector of the theory
while there is only little work on the gauge sector
(for a list of references see \cite{Gmeiner:2005vz}).
In one particularly interesting study \cite{schellekens} Dijkstra et. al
analyzed a huge ensemble of Gepner models to find possible Standard Model
candidates.

In this article we will describe the results of a survey of
the distribution of gauge sector observables and their
correlations for
supersymmetric intersecting D-brane models on a
$T^6/(\bZ_2\times\bZ_2)$ orientifold
background \cite{t6z2}.
In earlier work \cite{Blumenhagen:2004xx} these distributions were
estimated by a saddle point approximation,
later results have been obtained using a brute force computer analysis
\cite{Gmeiner:2005vz}.
This paper is organized as follows. In section \ref{sec_models}
we will briefly sum up the basic properties of the models we
are considering. In section \ref{sec_methods} we will describe
the algorithm we used to compute the results and in section
\ref{sec_results}
we will present the various results from the computer analysis.
Finally we will give some conclusions and an outlook to further directions
of research in section \ref{sec_outlook}.

\section{Models}\label{sec_models}
\subsection{Type II orientifold models}
The class of models we are considering are
supersymmetric
type IIA orientifold compactifications.
The orientifold projection
$\Omega$ combined with an antiholomorphic involution $\bar{\sigma}$,
which we will take to be complex conjugation in the following,
introduces orientifold O6-planes in the background, which carry
tension and negative RR-charge. To cancel the induced tadpoles one
introduces D6-branes, which is also very desirable
from a phenomenological point of view because they will provide
us with a realization of low-energy particle physics.
Both orientifold planes and D-branes, wrap Lagrangian
three-cycles $\pi_a$ in the internal manifold, which we take to be special
Lagrangian in order to preserve half of the supersymmetry. The
orientifold images of these cycles will be denoted by $\pi'_a$.

The homology group $H_3(M,\bZ)$ of three cycles in the compact manifold
$M$ splits under the action of $\Omega\bar{\sigma}$ into an even
and an odd part such that the only non-vanishing intersections are
between odd and even cycles. We can therefore choose a symplectic
basis $(\alpha_I,\beta_I)$ and expand $\pi_a$,$\pi'_a$ and $\pi_{O6}$ as
\be{eq_base}
\pi_a = \sum_{I=1}^{b_3/2}\left(X_a^I\alpha_I+Y_a^I\beta_I\right),\quad\quad
\pi'_a = \sum_{I=1}^{b_3/2}\left(X_a^I\alpha_I-Y_a^I\beta_I\right),\quad\quad
\pi_{O6} = \frac{1}{2}\sum_{I=1}^{b_3/2}L^I\alpha_I.
\ee
Chiral matter arises at the intersection of branes wrapping different
three-cycles. Generically we get bifundamental representations $(N_a,N_b)$
for two stacks with $N_a$ and $N_b$ branes. In addition we get matter
transforming in symmetric or antisymmetric representations of the
gauge group for each individual stack.
The multiplicities of these representations are given by the intersection
numbers of the three-cycles,
\be{eq_isn}
I_{ab} := \pi_a\circ \pi_b
        = \sum_{I=1}^{b_3/2}\left(X_a^IY_b^I-X_b^IY_a^I\right).
\ee
The possible representations are summarized in table \ref{tab_reps}.
\begin{table}[h]
\renewcommand{\arraystretch}{1.3}
%\tabsidecaption
\begin{tabular}{|c|c||c|c|}
\hline
reps. & multiplicity & reps. & multiplicity \\\hline
$(\N_a,\ov{\N}_b)$ & $I_{ab}$ & $\Sym_a$ & $\frac{1}{2}(I_{aa'} - I_{a{\rm O}6})$ \\
$(\N_a,\N_b)$ & $I_{ab'}$ & $\Anti_a$ & $\frac{1}{2}(I_{aa'} + I_{a{\rm O}6})$ \\ 
\hline
\end{tabular}%
\caption{Multiplicities of the chiral spectrum.}
\label{tab_reps}
\end{table}\\
The tadpole cancellation condition for $k$ stacks of $N_a$ D-branes
is given by
\be{eq_tad}
\sum_{a=1}^k N_a(\pi_a +\pi'_a)=4\pi_{O6} \quad\quad\leadsto\quad\quad
\sum_{a=1}^kN_aX_a^I=L^I.
\ee
The supersymmetry conditions are given by the calibration condition for
the cycles and an additional constraint to exclude anti-branes,
\be{eq_susy1}
\Im (\Omega_3)|_{\pi_a}=0,\quad\quad
\Re (\Omega_3)|_{\pi_a}>0,
\ee
where $\Omega_3$ is the holomorphic 3-form.
Written in the symplectic basis these equations read
\be{eq_susy2}
\sum_{I=1}^{b_3/2}Y_a^If_I = 0,\quad\quad
\sum_{I=1}^{b_3/2}X_a^Iu_I > 0,\quad\quad\mbox{where}\quad
f_I:=\int_{\beta_I}\Omega_3,\quad\quad
u_I:=\int_{\alpha_I}\Omega_3.
\ee
In addition to the tadpole and supersymmetry conditions
there exists a further constraint from K-theory \cite{Uranga:2000xp}.
It is a consistency condition to assure that the
$\bZ_2$ valued K-theory charge of an $Sp(2)$ probe brane
is conserved. If this constraint is violated, the model
suffers from a global gauge anomaly.

\subsection{$T^6/(\bZ_2\times\bZ_2)$}
In the sequel we will focus on a specific orientifold construction, a
compactification on $T^6/(\bZ_2\times\bZ_2)$. We will consider
only a special class of three-cycles on this torus, namely ''factorizable''
cycles, which can be described by their wrapping numbers $(n_i,m_i)$
along the basic one-cycles $\pi_{2i-1}, \pi_{2i}$ of the three
two-tori $T^6=\Pi_{i=1}^3T_i^2$.
To preserve the
symmetry generated by $\Omega\bar{\sigma}$, only two different shapes
of tori are possible, which can be parametrized by $b_i\in\{0,1/2\}$
and transform as
\be{eq_cyctr}
  \Omega\bar{\sigma}: \left\{\begin{array}{rcl}
    \pi_{2i-1} &\to& \pi_{2i-1}-2b_i\pi_{2i}\\
    \pi_{2i} &\to& -\pi_{2i}
  \end{array}\right..
\ee
For convenience we will work with the combination
$\tilde{\pi}_{2i-1}=\pi_{2i-1}-b_i\pi_{2i}$ and modified
wrapping numbers $\tilde{m}_i=m_i+b_in_i$.
Furthermore we introduce a rescaling factor
$c:=\left(\Pi_{i=1}^3(1-b_i)\right)^{-1}$
to get integer-valued coefficients. These are explicitly given by ($i,j,k\in\{1,2,3\}$ cyclic)
\be{eq_xydef}
X^0=cn_1n_2n_3,\quad X^i=-cn_i\tilde{m}_j\tilde{m}_k,\quad
Y^0=c\tilde{m}_1\tilde{m}_2\tilde{m}_3,\quad Y^i=-c\tilde{m}_in_jn_k.
\ee
Using these conventions the intersection numbers can be written as
\be{eq_myisn}
I_{ab}=\frac{1}{c^2}\left(\vec{X}_a\vec{Y}_b-\vec{X}_b\vec{Y}_a\right)
\ee
and the tadpole cancellation and supersymmetry conditions read
\be{eq_mytadsusy}
\sum_{a=1}^kN_a\vec{X}_a=\vec{L},\quad\quad
\sum_{I=0}^3\frac{Y^I}{U_I}=0,\quad\quad\sum_{I=0}^3X^IU_I>0,
\ee
where we used that the value of the physical orientifold charge is $8$ in
our conventions
and we defined the vector $\vec{L}:=\left(8c,\{8/(1-b_i\}\right)^T$.
The complex structure moduli $U_I$ can be defined in terms
of the radii $(R_i^{(1)},R_I^{(2)})$ of the three tori as
\be{eq_udef}
U_0=R_1^{(1)}R_1^{(2)}R_1^{(3)},\quad\quad
U_i=R_1^{(i)}R_2^{(j)}R_2^{(k)},\quad i,j,k\in\{1,2,3\}\,\mbox{cyclic}.
\ee
Finally the K-theory constraints can be expressed as
\be{eq_myk}
\sum_{a=1}^kN_aY_a^0\in2\bZ,\quad\quad
\frac{1-b_i}{c}\sum_{a=1}^kN_aY_a^i\in2\bZ,\quad
  i\in\{1,2,3\}.
\ee

\section{Methods}\label{sec_methods}
To do a statistical analysis we need our ensemble of vacua to be as large
as possible. If feasible we would like to determine \emph{all} solutions
of the tadpole equations, satisfying the constraints
from supersymmetry (\ref{eq_mytadsusy}) and K-theory
(\ref{eq_myk}).
Unfortunately an analytic solution to the tadpole equations, which are of
Diophantine type, is not possible. Even worse, it can be shown that the
problem of finding solutions falls in the class of NP complete problems
\cite{gareyjohnson1979}. This means basically that there is no way to find an
algorithm which generates solutions in polynomial time.

\subsection{The algorithm}\label{ssec_alg}
To develop an efficient algorithm solving the constraining equations, we
split the problem in three parts. In a first step possible values for the
wrapping numbers $X^I$ and $Y^I$ are computed, which fulfill the inequality
\be{eq_alg1}
  0 < \sum_{I=0}^3 X^I\, {U}_I \le \sum_{I=0}^3 L^I\, {U}_I
\ee
that can be derived using (\ref{eq_mytadsusy}).
In a second step we apply a fast algorithm to find partitions of natural
numbers according to the following equation
\be{eq_alg2}
  \sum_{a=1}^kS_a=C \quad\mbox{with}\quad
  S_a:=\sum_{I}N_a {U}_IX_a^I\quad\mbox{and}\quad
  C:=\sum_I L^I {U}_I.
\ee
Finally the terms $S_a$ of this partition are factorized into values for
$N_a$ and $X_a^I$ using the set of $X^I$ generated in step 1. Only
configurations that fulfill the K-theory constraints (\ref{eq_myk}) are
taken into account.
\begin{figure}[h]
%\sidecaption
\includegraphics[width=0.5\linewidth]{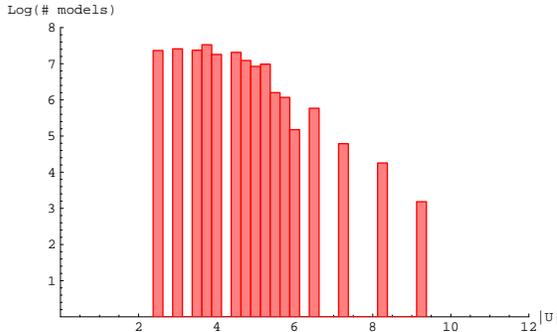}%
\caption{Number of solutions found depending on the absolute value
of the complex structure variables $U_I$.}
\label{fig_numsol}
\end{figure}

Note that in this procedure we have to chose a fixed set of complex structures
$U_I$ from the very beginning. To find all solutions we have therefore to
loop over all possible values for them. This is exactly the point where the
bad scaling of our algorithm leads to trouble. The time to compute
solutions grows exponentially with $U_I$. Within a total computing
time of more than six months on a high performance computer cluster we
have only been able to compute solutions up to an absolute value for $U_I$ of
12. The only point that saves our day is the experimental fact that the number
of solutions decreases exponentially with the absolute value of the complex
structures, as can be seen in figure \ref{fig_numsol}.
Being interested in a \emph{statistical} statement for the full ensemble
of solutions we do not have to worry about the fact that we have missed
some solutions with higher complex structures, because their influence
on the distribution of observables is negligible.

\section{Results}\label{sec_results}
\begin{figure}[h]
\subfigure[Rank distribution]{\label{fig_rkdist}\includegraphics[width=0.5\linewidth]{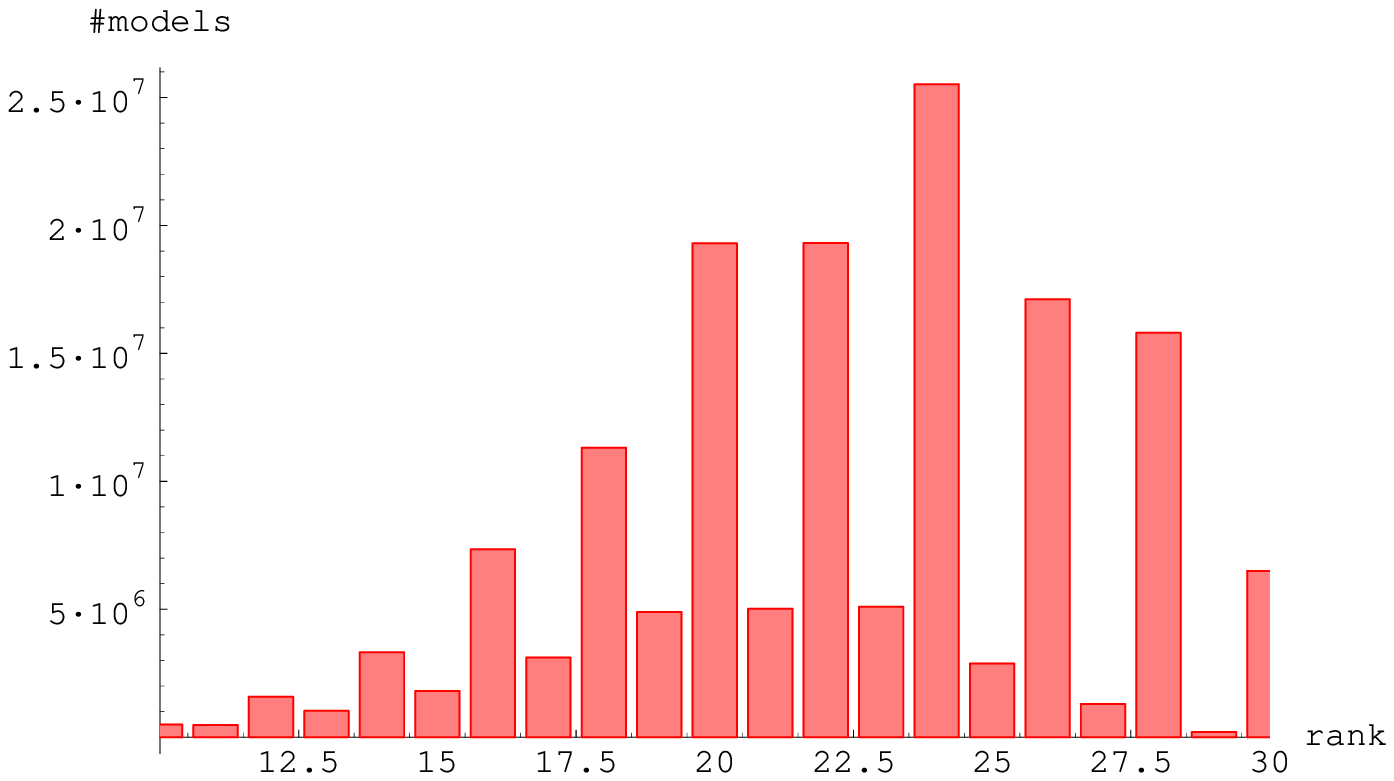}}%
\subfigure[$U(M)$ distribution]{\label{fig_undist}\includegraphics[width=0.5\linewidth]{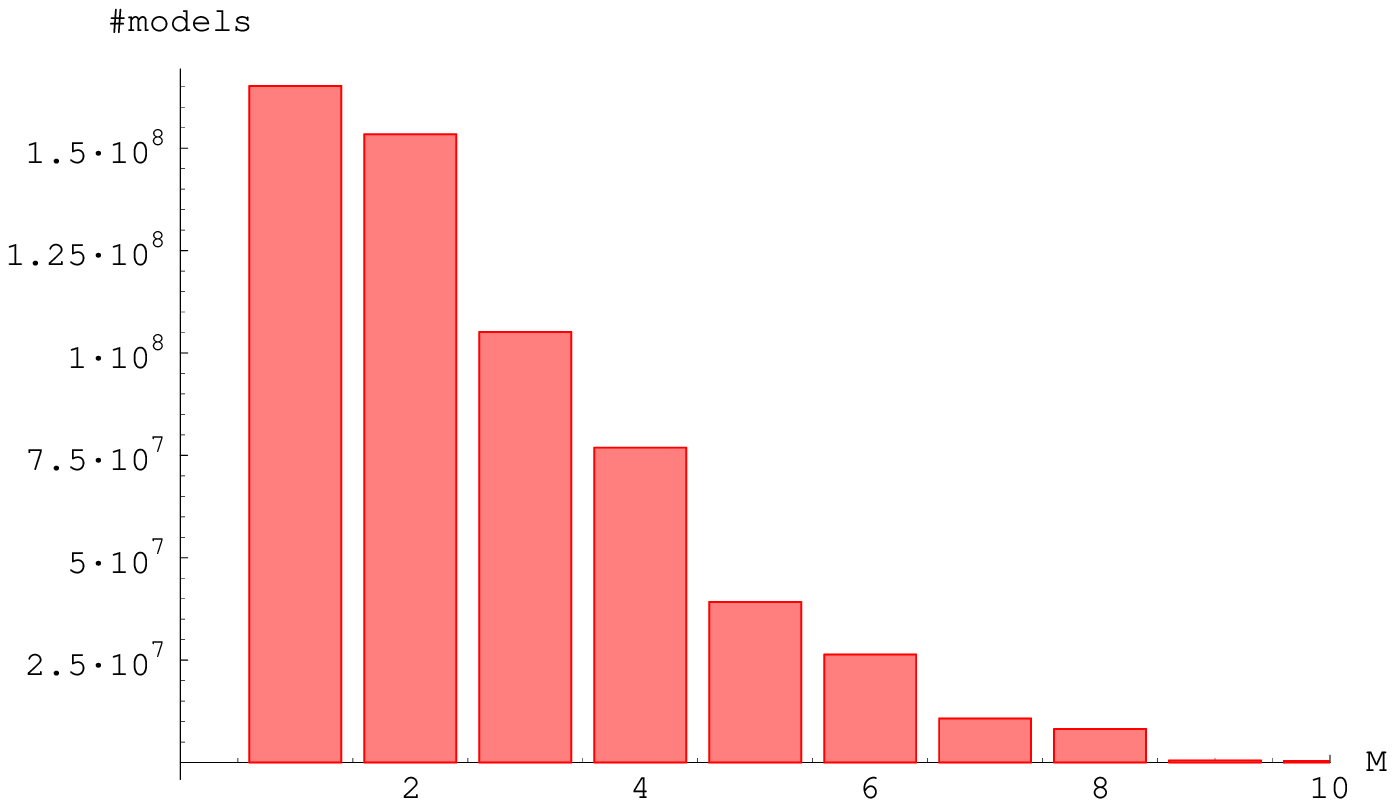}}%
\caption{Frequency distributions of total rank and $U(M)$ gauge groups of all
models.}
\label{fig_res1}
\end{figure}

Using the algorithm described in the last section we calculated a total number
of $\approx 1.6\times10^8$ models. Using this data we can now try to analyze
the observables of these models. The total rank $r$ of the gauge group for
example is given by $r=\sum_{a}N_a$. The resulting distribution is shown
in figure \ref{fig_rkdist}. Interesting is the suppression of odd values for
the total rank. This can be explained by the K-theory constraints
and the observation that the generic value for $Y^I$ is 0 or 1.
Therefore equation (\ref{eq_myk}) suppresses solutions with an odd value for
$r$. This suppression from the K-theory constraints is quite strong,
the number of solutions reduces by a factor of six compared to the situation
where these constraints are not enforced.

Another interesting quantity is the
distribution of $U(M)$ gauge groups, shown in figure \ref{fig_undist}.
We find that most models carry at least one $U(1)$ gauge group, corresponding
to a single brane, and stacks with a higher number of branes become more and
more unlikely. This could have been expected because small numbers occur
with a much higher frequency in the partition and factorization of natural
numbers.

\subsection{Standard Model-like configurations}\label{ssec_SM}
The special subset of models we are most interested in consists of those
which show some properties of the Standard Model or, to be more precise,
of the MSSM, since we are considering supersymmetric models only.
To realize the gauge group of the Standard Model we need generically four
stacks of branes (denoted by a,b,c,d) with two possible choices for the
gauge groups:
$U(3)_a \times U(2)_b \times U(1)_c \times U(1)_d$ or
$U(3)_a \times Sp(2)_b \times U(1)_c \times U(1)_d$.
To exclude exotic chiral matter from the first two factors we have to impose
the constraint that $\Sym_a$ and $\Sym_b$, the number of symmetric
representations, have to be zero. Models with only three stacks can also be
realized, but they suffer generically from having non-standard Yukawa
couplings.
Since we are not treating our models in so much detail and are more interested
in their generic distributions, we will include these three-stack constructions
in our analysis.

Another important ingredient for Standard Model-like configurations is the
existence of a massless $U(1)_Y$ hypercharge. This is in general a
combination $U(1)_Y=\sum_{i\in\{a,b,c,d\}}x_iU(1)_i$ of all four $U(1)$s such
that the condition $\sum_ix_iN_i\vec{Y}_i=0$ is fulfilled. More concretely
there are three possible ways to construct this hypercharge
$Q_Y^{(1)}=\frac{1}{6}Q_a+\frac{1}{2}Q_c+\frac{1}{2}Q_d$,~
$Q_Y^{(2)}=-\frac{1}{3}Q_a-\frac{1}{2}Q_b$~ or~
$Q_Y^{(3)}=-\frac{1}{3}Q_a-\frac{1}{2}Q_b+Q_d$,~
where choices 2 and 3 are only available for the first choice of gauge groups.
In total we have four ways to realize the Standard Model with massless
hypercharge\footnote{For a complete list of the realization of quarks and
leptons in the different setups please see \cite{Gmeiner:2005vz}, tables 2 and
3.}.

\begin{figure}[h]
%\sidecaption
\includegraphics[width=0.5\linewidth]{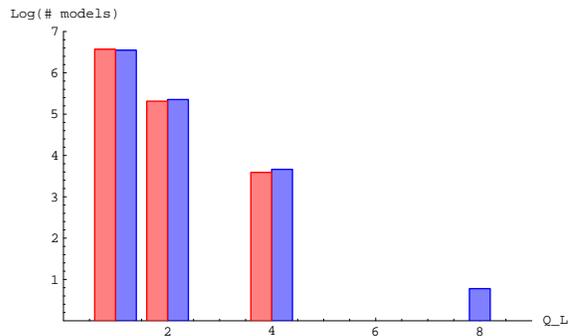}%
\caption{Number of quark and lepton generations with (red bars on the left) and
without (blue bars on the right) enforcing a massless $U(1)$ .}
\label{fig_gensm}
\end{figure}
The first question one would like to ask, after having defined what a ''Standard
Model'' is in our setup, concerns the frequency of such configurations in the
space of all solutions. Put differently: \emph{How many Standard Models
with three generations of quarks and leptons did we find?}. The answer to
this question is \emph{zero}, even if we relax our constraints and allow for
a massive hypercharge (which is rather fishy from a phenomenological point of
view). The result of the analysis can be seen 
in figure \ref{fig_gensm}. This is a rather strange result, since we know
that models with three families have already been constructed in our setup
(e.g. \cite{sm}).
A detailed analysis of the models in the literature shows that all models
which are known use (in our conventions) large values for the complex structure
variables $U_I$ and therefore did not appear in our analysis (see section
\ref{ssec_alg}). On the other hand we know that the number of models decreases
exponentially with higher values for the complex structures. Therefore we
conclude that Standard Models with three generations are highly suppressed in
this specific setup, more specifically the suppression factor can be estimated
to be about $10^{-9}$ (cf. \cite{Gmeiner:2005vz}, section 5.1).

\subsection{Hidden sector}\label{ssec_hidden}
\begin{figure}[h]
\subfigure[Rank distribution]{\label{fig_hrkdist}\includegraphics[width=0.5\linewidth]{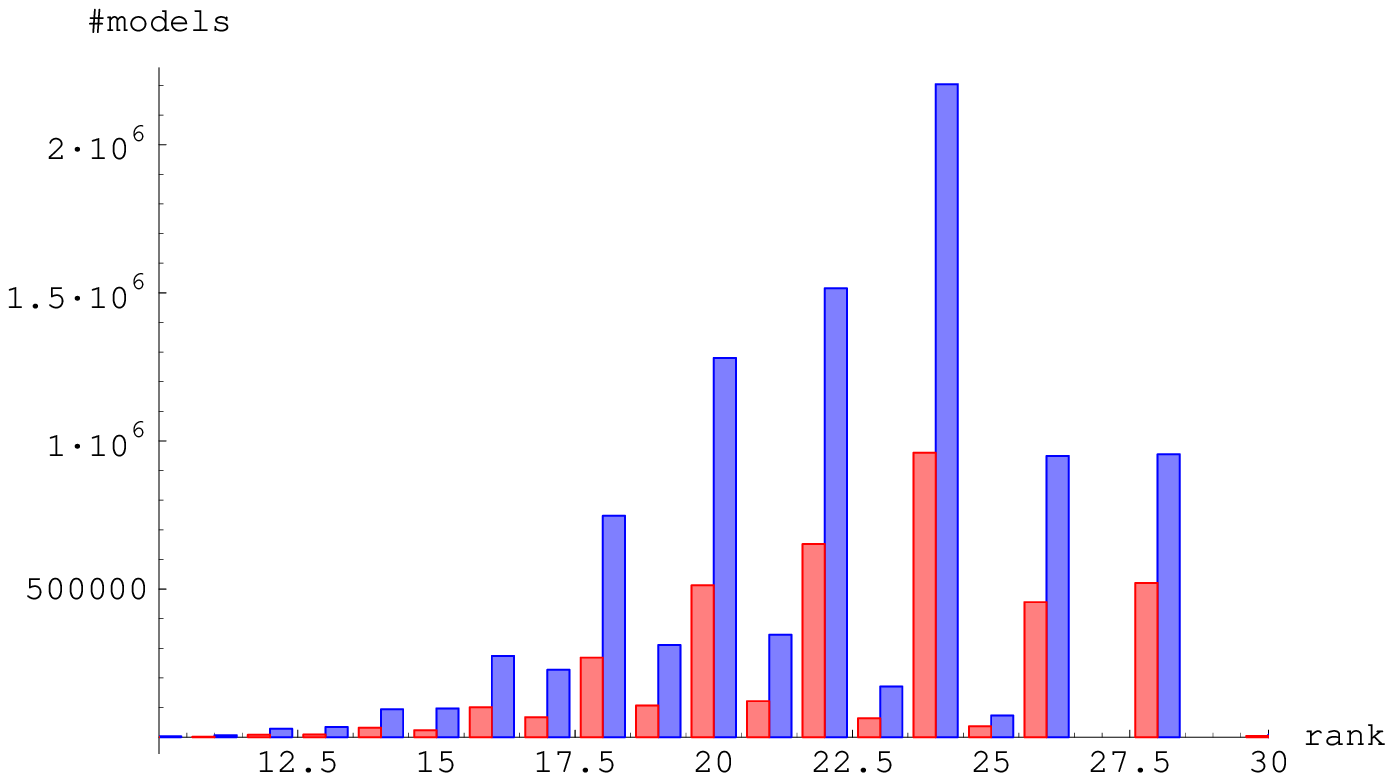}}%
\subfigure[$U(M)$ distribution]{\label{fig_hundist}\includegraphics[width=0.5\linewidth]{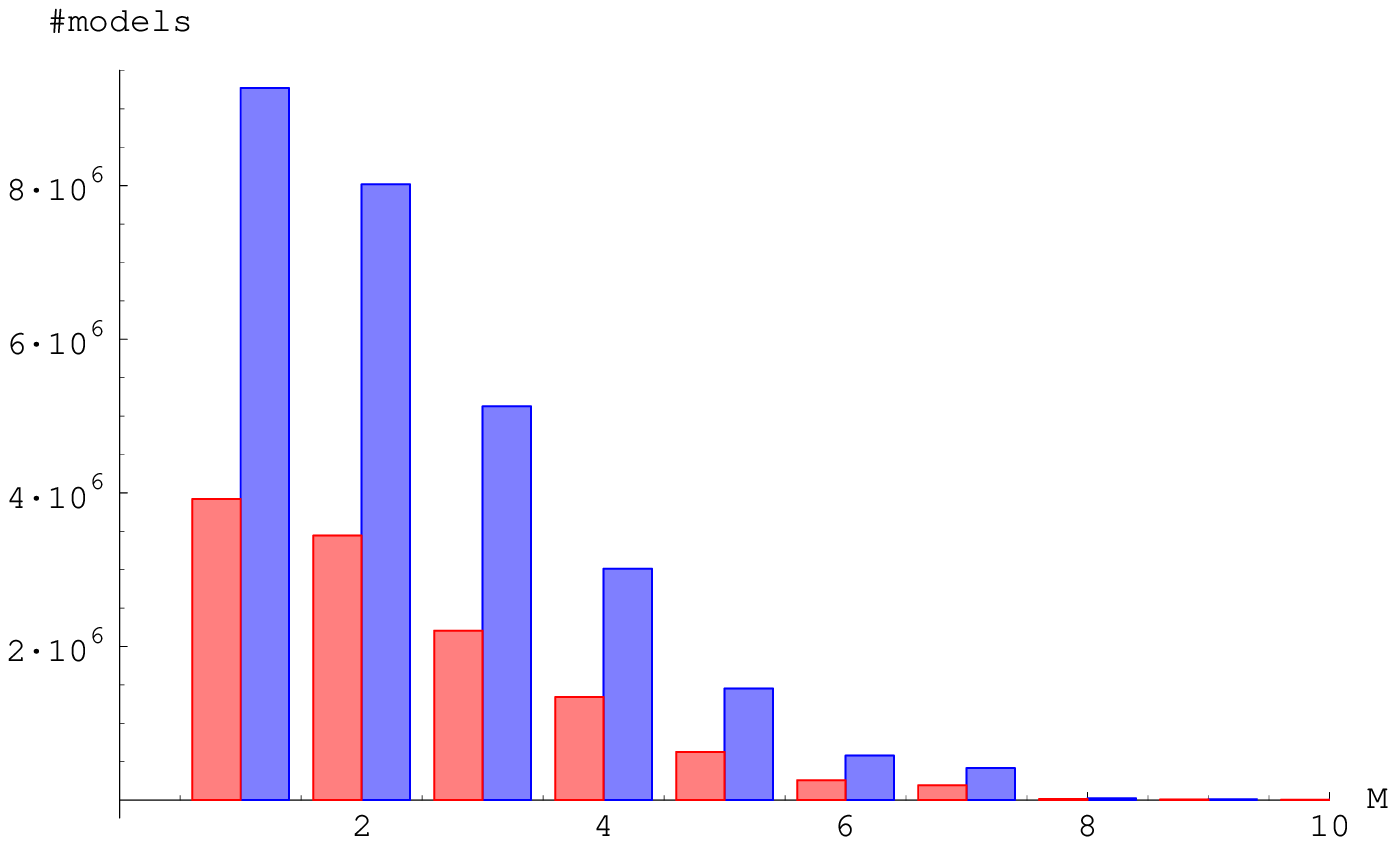}}%
\caption{Frequency distributions of total rank and $U(M)$ gauge groups in the
hidden sector of MSSM-models (red bars on the left) and MSSM models with
massive $U(1)$ (blue bars on the right).}
\label{fig_res2}
\end{figure}

Although we did not find a nice three generation Standard Model in our
analysis, it might still be interesting to ask questions about the hidden
sector of these models. As has been realized in \cite{Gmeiner:2005vz} many
of the properties of our models can be regarded to be independent of each
other, which means that the statistical analysis of the hidden sector of 
any model with specific visible gauge group would lead to very similar
results. This is indeed the case, as can be seen in figure \ref{fig_res2}.
Moreover, comparing this result with the distributions of the full set of
models (figure \ref{fig_res1}) shows that qualitatively the restriction to
a specific visible sector does not change the distribution of gauge group
observables.

The frequency distributions of the hidden sector might be compared
with a similar analysis for Gepner models carried out in \cite{schellekens}.
To do a direct comparison with their results, we have to restrict the number
of branes in the hidden sector to a maximum of three. The resulting
distribution turns out to be qualitatively
very similar to the Gepner model results, although our ensemble of models
is much smaller and not restricted to three generation models.

\subsection{Gauge couplings}\label{ssec_gcp}
The gauge sector observables considered in the last paragraphs all belong
to the topological sector of the theory, in the sense that they are defined
by the wrapping numbers of the branes and independent of the geometric
moduli. This does not apply to the gauge couplings, which explicitly
do depend on the complex structures, following the derivation in
\cite{Blumenhagen:2003jy}, which in our conventions reads
\be{eq_gcoup}
  \frac{1}{\alpha_a}=\frac{M_{Planck}}{2\sqrt{2}M_s\kappa_a}\,\frac{1}{c\sqrt{\prod_{i=1}^3R_1^{(i)}R_2^{(i)}}}\sum_{I=0}^{3}X^I U_I,
\ee
where $\kappa_a=1$ or $2$ for an $U(N)$ or $Sp(2N)$ stack respectively.

\begin{figure}[h]
\begin{minipage}{0.46\linewidth}
\includegraphics[width=\linewidth]{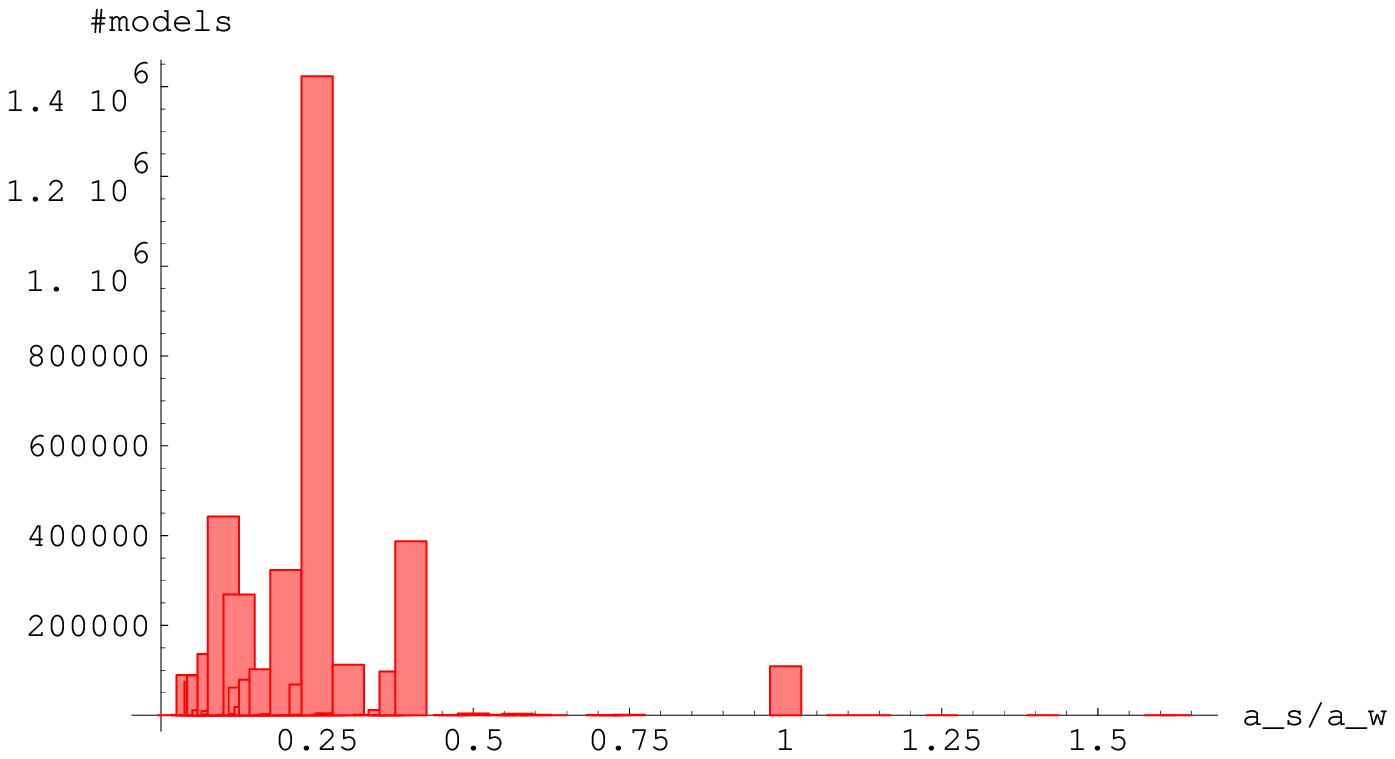}%
\caption{Frequency distribution of $\ga_s/\ga_w$ in Standard Model-like
configurations.}
\label{fig_gcdist}
\end{minipage}
\hfil
\begin{minipage}{0.5\linewidth}
\includegraphics[width=0.9\textwidth]{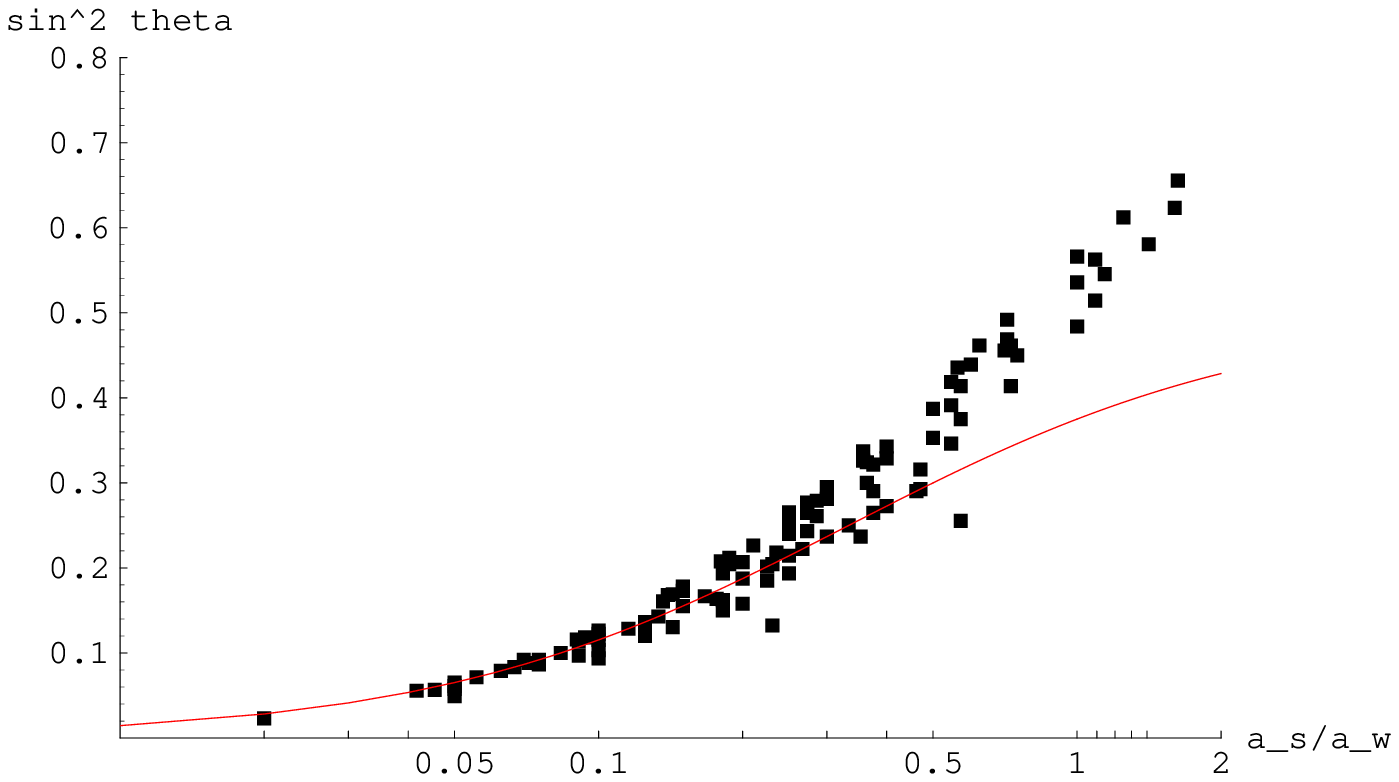}%
\caption{Values of $sin^2\theta$ depending on $\ga_s/\ga_w$.
Each dot represents a class of models with these values.}
\label{fig_sint}
\end{minipage}
\end{figure}

If one wants to perform an honest analysis of the coupling constants,
one would have to compute their values at low energies using the
renormalization group equations. We will not do this here, but instead
look at the distribution of $\ga_s/\ga_w$ at the string scale.
A value of one at the string scale does then not necessarily mean
unification at lower energies, but it could be taken as a hint in
this direction.
The result is shown in figure \ref{fig_gcdist} and it turns out that only
2.75\% of all models actually do show gauge unification at the string
scale.

Furthermore we would like to analyze the value of the
Weinberg angle $\sin^2\theta=\ga_Y/(\ga_Y+\ga_w)$ depending on the ratio
$\ga_s/\ga_w$ and check the following 
relation between the three couplings, which was proposed in
\cite{Blumenhagen:2003jy} and should hold for a large class
of intersecting brane models
\be{eq_crel}
  \frac{1}{\ga_Y}=\frac{2}{3}\frac{1}{\ga_s}+\frac{1}{\ga_w}\quad\leadsto\quad
  \sin^2\theta=\frac{3}{2}\,\frac{1}{\ga_w/\ga_s+3}.
\ee
The result is shown in figure \ref{fig_sint}, where we included a red line
that represents the relation (\ref{eq_crel}). The fact that actually 88\% of
all models obey this relation is a bit obscured by the plot, because each dot
represents a class of models and small values for $\ga_s/\ga_w$ are highly
preferred (cf. figure \ref{fig_gcdist}).

Comparing these results with the Gepner model analysis of \cite{schellekens}
we find, in contrast to the case of hidden sector gauge groups (see section
\ref{ssec_hidden}), no agreement. The fraction of models obeying
(\ref{eq_crel}) was found to be only about 10\% in the Gepner model case.
This might be traced back to the observation that in contrast to the
topological
data of gauge groups we are dealing with geometrical aspects here, which
are more influenced by the fact that we are working in a large radius regime,
while the Gepner model analysis is done at small radius.

\subsection{Correlations}\label{ssec_corr}
\begin{figure}[h]
%\sidecaption
\includegraphics[width=0.4\linewidth,trim=0mm 10mm 0mm 10mm,clip]{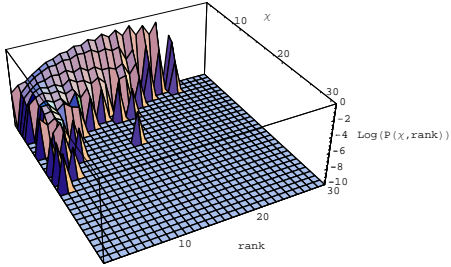}%
\caption{Frequency distribution of models with Standard Model gauge group
depending on total rank and chirality.}
\label{fig_corr}
\end{figure}

An interesting question that we raised in the introduction concerns the
correlation of observables.
If different properties of our models were correlated,
independently of the specific visible gauge group, this would provide us
with some information about the generic behavior of this class of models.
As it turns out, this is indeed the case. A nice example is the correlation
between the total rank of the gauge group and the ''mean chirality'' of
a model. The mean chirality is a quantity defined as
$\chi := n\sum_{a\neq b}(I_{a'b}-I_{ab})$, where $n$ is some normalization.
The difference between the two intersection numbers computes the amount of
chiral matter in a bifundamental representation between the stacks $a$ and
$b$. $\chi$ measures therefore the amount of chirality that is present in a
specific model.
Looking at the distribution of models for specific values of rank and chirality
we find that their values are not independent.
Moreover, this behavior is generic and the shape of figure \ref{fig_corr}
does not change if we require a different visible sector gauge group.

\section{Conclusions}\label{sec_outlook}
We presented some aspects of an explicit statistical analysis of a specific
construction in the class of type II orientifold models.
An interesting result is that models with three generations of Standard Model
particles are highly suppressed in this setup. We found that the observables
in the hidden sector are independent of the specific gauge group in the
visible sector. Furthermore we found evidence for a correlation between
observables. Comparing our results to the small radius analysis of Gepner
models in \cite{schellekens} we find agreement for topological quantities, but
disagreement for more geometrical properties like the gauge couplings, as
could have been expected.

It would be very interesting to compare these results with statistical data
of other constructions, like heterotic compactifications. Equally desirable
would be the inclusion of fluxes in our statistical treatment and the analysis
of grand unified gauge groups like $SU(5)$ in the visible sector.

\begin{acknowledgement}
I would like to thank the organizers of the RTN network conference
\emph{Constituents, Fundamental Forces and Symmetries of the Universe}
in Corfu, Greece for the invitation.
I am grateful to Ralph Blumenhagen, Gabriele Honecker, Dieter
L\"ust and Timo Weigand for collaboration on this project.
\end{acknowledgement}

\end{document}